# Capacitive Spring Softening in Single-Walled Carbon Nanotube Nanoelectromechanical Resonators

Chung Chiang Wu and Zhaohui Zhong[*]

Department of Electrical Engineering and Computer Science, University of Michigan, Ann Arbor, MI 48109

[*]To whom all correspondence should be addressed. zzhong@umich.edu

**ABSTRACT**

We report the capacitive spring softening effect observed in single-walled carbon nanotube (SWNT) nanoelectromechanical (NEM) resonators. The nanotube resonators adopt dual-gate configuration with both bottom-gate and side-gate capable of tuning the resonance frequency through capacitive coupling. Interestingly, downward resonance frequency shifting is observed with increasing side-gate voltage, which can be attributed to the capacitive softening of spring constant. Furthermore, in-plane vibrational modes exhibit much stronger spring softening effect than out-of-plan modes. Our dual-gate design should enable the differentiation between these two types of vibrational modes, and open up new possibility for nonlinear operation of nanotube resonators.

**Keywords:** carbon nanotube, nanoelectromechanical resonators, nonlinearity,

Due to their low mass density and high Young's modulus [1-4], SWNTs offer great promise as ultrahigh frequency NEM resonators with applications in ultrasmall mass and force sensing [5-10]. To detect the mechanical vibration of nanotube resonators, various methods have been explored so far [7, 11-13]. In particular, nanotube resonators can be actuated and detected simultaneously through electrostatic gate coupling, offering *in situ* frequency tuning over wide frequency range [7-8]. The gate induced frequency tuning of NEM resonators is known to be governed by two mechanisms: the elastic hardening effect which increases the resonance frequencies, and the capacitive softening effect which decreases the resonance frequencies [14-17]. While the elastic hardening is originated from the increased beam tension, the capacitive softening is caused by beam oscillating in a constant electric field, which reduces the effective spring constant. Although elastic hardening effect has been widely reported in SWNT NEM resonators [7-8, 18], the field-induced capacitive spring softening has rarely been observed [19]. In a conventional nanotube resonator design [7, 13, 18], only bottom-gate is used for frequency tuning. The nanotube vibrational motions [19-20] are perpendicular to the electric field direction, resulting in negligible spring softening. To this end, we demonstrate SWNT resonators with a dual-gate configuration, which enables both upward and downward frequency tuning by exploring elastic hardening and capacitive softening effects. Further, we discover that in-plane vibrational modes exhibit much stronger spring softening than out-of-plan mode, suggesting that the dual-gate technique can serve as an experimental method in differentiating vibrational modes.

SWNT NEM resonators with a dual-gate configuration are fabricated using the one-step direct transfer technique discussed previously [21]. Briefly, suspended SWNTs are grown across pillars on a transparent quartz substrate by the chemical vapor deposition (CVD) method [22-24], while pre-designed electrodes are fabricated on a separate device substrate by conventional lithography. The transfer of suspended SWNTs to the device substrate is implemented by simply bringing two substrates into contact. Fig. 1(a) shows the scanning electron microscope (SEM) image of a typical dual-gate SWNT resonator. Suspended SWNT (indicated by the arrow) spans across the source (S) and drain (D) electrodes (in green) with an underneath bottom-gate (BG) and nearby side-gates (SG) electrostatically coupled to the nanotube. For a typical device, the S and D electrodes are 2 μm wide, separated by 3 μm, and the distance between nanotube and the BG is 1 μm. To explain how our dual-gate nanotube resonators can realize frequency-tuning through both elastic hardening and capacitive softening mechanisms, a qualitative sketch is illustrated in Fig. 1(b). SWNT resonators exhibit two types of vibration modes [19-20]: the in-plane mode (left-top panel) moves along the y-direction; and the out-of-plane mode (right-top panel), vibrating like a jumping rope, moves along the x-direction. When a voltage $V_{bg}$ is applied on the BG electrode, the electrostatic force will pull down the nanotube toward the gate (gray arrow, along the z-direction), thus increasing the nanotube tension and resulting in elastic hardening. Since the electrostatic force is perpendicular to the vibration directions of both in-plane and out-of-plane modes, the effect of capacitive softening is negligible. On the other hand, when a side-gate voltage $V_{sg}$ is applied, the electrostatic force (red arrow) will have component along y-direction in addition to z-direction. As the result, the in-plane vibrational mode will be impeded by the electrostatic force, leading to a strong capacitive softening effect compared to out-of-plane mode.

To experimentally examine these two frequency tuning mechanisms, we systematically applied voltages on both bottom-gate and side-gate electrodes. A small AC driving voltage of 10 mV applied to the drain electrode was used to actuate the resonator through electrostatic interaction, and resonance frequency was detected by measuring the frequency-dependent mixing current $I_{mix}$ described in previous literatures [7-8]. We first studied the frequency tuning using bottom-gate electrode. As shown in Fig. 2(a) inset, nanotube resonances are clearly visible, and resonance frequency increases from 11 MHz to 22 MHz as $|V_{bg}|$ increases from 0 V to 2 V. We further extract resonance frequencies at different $V_{bg}$ between -2V to 2V, and the results can be fitted with a parabolic function [Fig. 2(a)]. Our observation of frequency tuning using BG electrode agrees well with elastic hardening effect reported on nanotube resonators [7, 18]. At small BG voltage, the nanotube resonators operate in the bending regime [7, 18-20], in which resonance frequency $f$ depends quadratically on the BG voltage [3, 19-20]:

$$f = f_0 + 0.28 \frac{C'}{\sqrt{96s}} \sqrt{\frac{1}{\mu EI}} V_{bg}^2 = f_0 + A V_{bg}^2$$

(1)

where $f_0$ is the fundamental frequency, $C'$ is the first derivative of capacitance to nanotube/gate distance, $s$ is the slack, $\mu$ is linear mass density, $E$ is Young's modulus, $I$ is the moment of inertia, and $A$ is termed as elastic hardening tuning coefficient. Fitting the experimental data in Fig. 2(a) with equation (1) yields measured coefficient $A$ of $1.9 \times 10^6$ HzV$^{-2}$. To compare $A$ with the theoretical value $A_{theory}$, we adopted a cylinder over a metal plane to model the capacitance $C = \frac{2\pi\varepsilon_0 L}{\ln(\frac{2Z}{d})}$, where $\varepsilon_0$ is the dielectric constant; $L$ is the nanotube length; $Z$ is the separation between tube and bottom electrode; $d$ is the diameter of nanotube. Assuming a small slack of 1% and typical SWNT parameters [19], the $A_{theory}$ value calculated using equation (1) is $2.5 \times 10^6$ HzV$^{-2}$, which agrees well with measured $A$.

Next, we examined the frequency dependence of side-gate voltage [Fig. 2(b)]. Interestingly, as shown in Fig. 2(b) inset, nanotube resonance frequency decreases as $|V_{sg}|$ increases. The extracted resonance frequencies at different $V_{sg}$ between -10V to 10V exhibit a negative curvature with increasing field. The observed downward frequency tuning by SG is in strong contrast to elastic hardening, but it can be explained by the capacitive softening effect. The frequency dependence of the capacitive softening effect can be expressed as [14-15]:

$$f^2 = f_0^2 - \frac{C''V_{sg}^2}{8\mu L \pi^2} = f_0^2 - BV_{sg}^2$$

(2)

where $C''$ is the second derivative of capacitance, and $B$ is termed the capacitive softening coefficient. Again, fitting the experimental data in Fig. 2(b) with equation (2) yields the measured coefficient $B$ of $0.8 \times 10^{12}\,\text{Hz}^2\text{V}^{-2}$. To calculate the theoretical value of $B_{theory}$, a factor $\gamma$ is introduced into the capacitance to compensate the offset geometry of SG, $C = \frac{2\pi\varepsilon_0 L \gamma}{\log(\frac{2Z}{d})}$. By comparing the $G$-$V_g$ transfer curves for BG and SG, we estimate a $\gamma \sim 1/3$. Using typical SWNT parameters, we calculate a $B_{theory}$ from equation (2) of $1.1 \times 10^{12}\,\text{Hz}^2\text{V}^{-2}$, which agrees well with measured $B$. Similar results have been observed on 3 dual-gate resonators, offering a reliable approach for studying both elastic hardening and capacitive softening effects in nanotube resonators for the first time.

SWNT resonators are known to exhibit multiple vibrational states, including in-plane, out-of-plane, and their higher order modes [19-20]. We therefore examined the capacitive softening effect for different vibrational modes. Figure 3(a) shows the mixing current (in color) plotted as a function of driving frequency and SG voltage. Three resonance modes are clearly visible with $f$ up to 23MHz, and they all exhibit capacitive softening effect with applied $V_{sg}$. The $V_{sg}$-dependent resonance frequencies for all three modes are plotted in the inset of Figure 3(a). Importantly, three vibrational modes show drastically different tunability. Fitting the experimental data in Figure 3(a) inset with equation (2) yields $B$ = 0.16, 0.8, and 0.76×10$^{12}$Hz$^2$V$^{-2}$ for vibrational modes from bottom to top, respectively. The two higher frequency modes exhibit ~5 times larger capacitive softening effect compared to the lowest frequency mode. Similar measurements were performed on 3 other SWNT resonators with the same device geometry, and the results are shown in Fig 3(b). The softening coefficients for the first vibration modes are always much smaller than those of higher order modes, differed by 4 to 6.3 times. This observation agrees with our qualitative analysis shown in Fig. 1(b), where differences in softening coefficients are expected for in-plane and out-of-plane modes. Therefore, we attribute the first vibrational mode with much smaller softening coefficient as out-of-plane mode, and the higher order vibrational modes with larger softening coefficients as in-plane modes. Our results also agree with the theoretical prediction of the first vibrational mode being the fundamental out-of-plane mode [19-20].

Last, we investigated how the coupling of BG and SG affects the capacitive softening effect. The $V_{sg}$-dependent resonance frequencies for the second vibrational mode shown in Fig. 3(a) are plotted for different fixed $V_{bg}$ [Fig. 3(c)]. As $|V_{bg}|$ increases from 0V to 2V, the downward frequency tuning is still clearly visible, but the curves are shifted toward higher $V_{sg}$ value. The shift rises from our dual-gate geometry, where the charge neutral point will shift as voltage being applied onto the BG. The capacitive softening equation can be modified by including an offset voltage, $V_0$, to account for the effect of BG:

$$f^2 = f_0^2 - \frac{C''(V_{gate} - V_0)^2}{8\mu L \pi^2} = f_0^2 - B(V_{gate} - V_0)^2$$

(3)

Fitting data in Fig. 3(c) with equation (3), we extracted the softening coefficient $B$ and offset voltage $V_0$, and the results are plotted in Fig 3(d). The softening coefficient (red circles) remains nearly constant at different $V_{bg}$ voltage, while $V_0$ (blue squares) varies linearly with respect to $V_{bg}$. A linear fit of $V_0$ vs. $V_{bg}$ yields a slope of ~3, suggesting that the BG is about three times more effective than SG for electrostatic charging.

In summary, we report the observation of capacitive softening effect in SWNT NEM resonators with dual-gate configuration. While in-plane vibrational modes show strong softening effect when SG voltage is applied, the fundamental out-of-plan mode exhibits small/negligible spring constant softening. Our results not only provide a new experimental tool for differentiating nanotube vibrational modes, but also enable additional freedom for exploring non-linear effects in nanotube resonators.

**Acknowledge**

The work is supported by the startup fund provide by the University of Michigan. This work used the Lurie Nanofabrication Facility at University of Michigan, a member of the National Nanotechnology Infrastructure Network funded by the National Science Foundation.

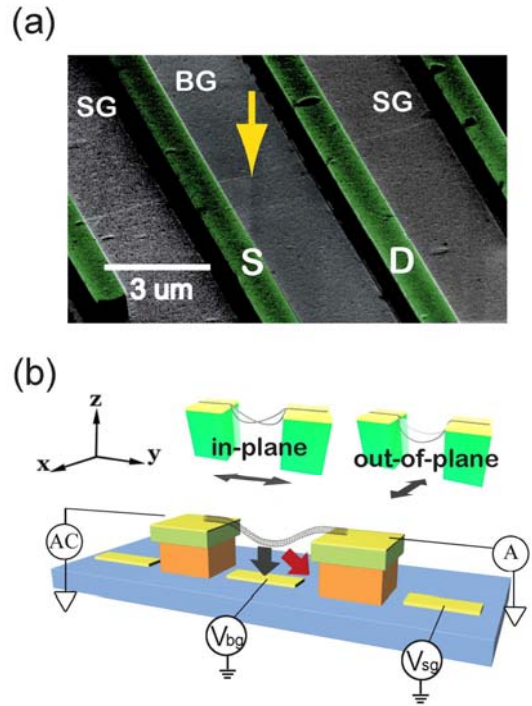

FIG. 1. (Color online). (a) The SEM image of SWNT NEM resonators. The source and drain electrodes are colorized (green) and 50-nm-thick Au is used as the bottom-contact metal. The arrow indicates the position of a suspended SWNT. (b) A qualitative sketch illustrates how electrostatic force interacts with different resonance modes when a bias voltage is applied on BG or SG electrodes.

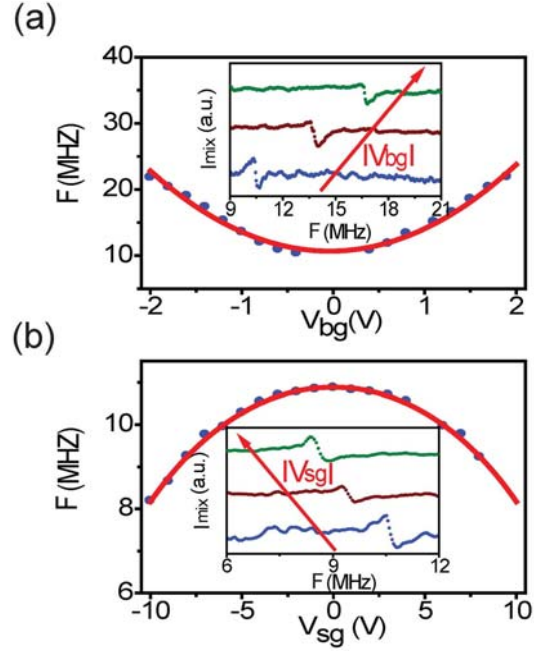

FIG. 2. (Color online). (a) Elastic hardening effect, and (b) Capacitive softening effect, observed on a SWNT NEM resonator by biasing BG and SG, respectively. The single-source mixing technique is used for resonance actuation and detection with $\delta V_{sd}$ = 10 mV, and measurement is done in a vacuum chamber at pressure below $10^{-4}$ torr. Insets of (a) and (b): mixing current ($I_{mix}$) vs. driving frequency ($f$) at different BG and SG voltages. Resonance peak shifts to higher frequency as $|V_{bg}|$ = 0.5, 1, and 1.5 V is applied, and shift to lower frequency as $|V_{sg}|$ = 3, 7, and 9 V is applied.

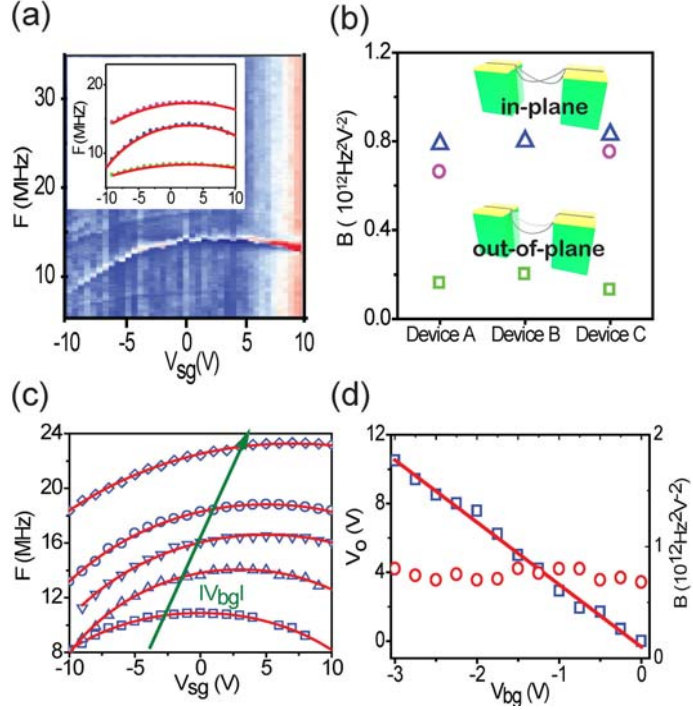

FIG. 3. (Color online) (a) Mixing current $I_{mix}$ (color scale) is plotted as a function of driving frequency and $V_{sg}$. Inset: Resonance peaks (blue dots) of different modes are plotted with respect to $V_{sg}$. Red lines are fitting curves with equation (3). (b) Measured softening coefficients ($Bs$') of different modes for three resonators. The softening coefficients for the first vibration modes (green squares) are always much smaller than those of higher order modes (blue triangles and purple circles), differed by 4 to 6.3 times. (c) Resonance frequencies (blue dots) vs. $V_{sg}$ curves for $V_{bg}$ = 0, -1, -1.25, -1.5, and -2V, from bottom to top. The neutral point $V_0$ shifts to higher voltages as $|V_{bg}|$ increases. (d) Offset voltage $V_0$ (blue squares) and coefficient $B$ (red circles) vs. $V_{bg}$ extracted from (c). A linear fit of $V_0$ vs. $V_{bg}$ yields a slope of ~3.